%% file: main.tex
\pdfoutput=1
\documentclass[12pt,a4paper]{article}

\usepackage{ifthen} % for conditional statements
\newboolean{pdflatex}
\setboolean{pdflatex}{true} % False for eps figures 

\newboolean{articletitles}
\setboolean{articletitles}{true} % False removes titles in references

\newboolean{uprightparticles}
\setboolean{uprightparticles}{false} %True for upright particle symbols

\input{preamble}
\begin{document}

%%%%%%%%%%%%%%%%%%%%%%%%%
%%%%% Title     %%%%%%%%%
%%%%%%%%%%%%%%%%%%%%%%%%%
\renewcommand{\thefootnote}{\fnsymbol{footnote}}
\setcounter{footnote}{1}
\input{title-LHCb-PAPER}

\renewcommand{\thefootnote}{\arabic{footnote}}
\setcounter{footnote}{0}

%%%%%%%%%%%%%%%%%%%%%%%%%
%%%%% Main text %%%%%%%%%
%%%%%%%%%%%%%%%%%%%%%%%%%
\pagestyle{plain} % restore page numbers for the main text
\setcounter{page}{1}
\pagenumbering{arabic}

% %%%%%%% CHOOSE --------
%% ----------------------------------
%% Line numbering on the left margin 
%% ----------------------------------
%% Uncomment during review phase. 
%% Comment it out before a final submission.
%%\linenumbers
%% --------------------------------
% %%%%%%%%%%%%% ---------

\input{body}
\input{acknowledgements}

\bibliographystyle{LHCb}

\ifx\mcitethebibliography\mciteundefinedmacro
\PackageError{LHCb.bst}{mciteplus.sty has not been loaded}
{This bibstyle requires the use of the mciteplus package.}\fi
\providecommand{\href}[2]{#2}

\input{appendix}

\end{document}

%% file: preamble.tex
\usepackage{microtype}
\usepackage{lineno}  % for line numbering during review
\usepackage{xspace} % To avoid problems with missing or double spaces after
\usepackage{graphicx}  % to include figures (can also use other packages)
\usepackage{color}
\usepackage{colortbl}
\graphicspath{{./figs/}} % Make Latex search fig subdir for figures

\usepackage{amsmath} % Adds a large collection of math symbols
\usepackage{amssymb}
\usepackage{amsfonts}
\usepackage{bm}
\usepackage{upgreek} % Adds in support for greek letters in roman typeset

\newcommand*\patchAmsMathEnvironmentForLineno[1]{%
\expandafter\let\csname old#1\expandafter\endcsname\csname #1\endcsname
\expandafter\let\csname oldend#1\expandafter\endcsname\csname
end#1\endcsname
 \renewenvironment{#1}%
   {\linenomath\csname old#1\endcsname}%
   {\csname oldend#1\endcsname\endlinenomath}%
}
\newcommand*\patchBothAmsMathEnvironmentsForLineno[1]{%
  \patchAmsMathEnvironmentForLineno{#1}%
  \patchAmsMathEnvironmentForLineno{#1*}%
}
\AtBeginDocument{%
\patchBothAmsMathEnvironmentsForLineno{equation}%
\patchBothAmsMathEnvironmentsForLineno{align}%
\patchBothAmsMathEnvironmentsForLineno{flalign}%
\patchBothAmsMathEnvironmentsForLineno{alignat}%
\patchBothAmsMathEnvironmentsForLineno{gather}%
\patchBothAmsMathEnvironmentsForLineno{multline}%
}

\usepackage{hyperref}    % Hyperlinks in references
\usepackage[all]{hypcap} % Internal hyperlinks to floats.

\input{lhcb-symbols-def} % Add in the predefined LHCb symbols

\usepackage{cite} % Allows for ranges in citations
\usepackage{mciteplus}

%% file: lhcb-symbols-def.tex
\def\lhcb {\mbox{LHCb}\xspace}
\def\ux85 {\mbox{UX85}\xspace}

\def\babar  {\mbox{BaBar}\xspace}

\ifthenelse{\boolean{uprightparticles}}%
{

 \def\PDelta      {\ensuremath{\Delta}\xspace}                 
 \def\PXi      {\ensuremath{\Xi}\xspace}                 
 \def\PLambda      {\ensuremath{\Lambda}\xspace}                 
 \def\PSigma      {\ensuremath{\Sigma}\xspace}                 
 \def\POmega      {\ensuremath{\Omega}\xspace}                 
 \def\PUpsilon      {\ensuremath{\Upsilon}\xspace}                 
 
 \def\PB      {\ensuremath{\mathrm{B}}\xspace}                 
                  
 \def\PD      {\ensuremath{\mathrm{D}}\xspace}

 \def\PK      {\ensuremath{\mathrm{K}}\xspace}

 \def\Pb      {\ensuremath{\mathrm{b}}\xspace}                 
 \def\Pc      {\ensuremath{\mathrm{c}}\xspace}

 \def\Pi      {\ensuremath{\mathrm{i}}\xspace}

 \def\Ps      {\ensuremath{\mathrm{s}}\xspace}

}
{

 \mathchardef\PDelta="7101
 \mathchardef\PXi="7104
 \mathchardef\PLambda="7103
 \mathchardef\PSigma="7106
 \mathchardef\POmega="710A
 \mathchardef\PUpsilon="7107
                  
 \def\PB      {\ensuremath{B}\xspace}                 
                  
 \def\PD      {\ensuremath{D}\xspace}

 \def\PK      {\ensuremath{K}\xspace}

 \def\Pb      {\ensuremath{b}\xspace}                 
 \def\Pc      {\ensuremath{c}\xspace}

 \def\Pi      {\ensuremath{i}\xspace}

 \def\Ps      {\ensuremath{s}\xspace}

}

\def\squark    {\ensuremath{\Ps}\xspace}

\def\cquark    {\ensuremath{\Pc}\xspace}

\def\bquark    {\ensuremath{\Pb}\xspace}

\def\kaon  {\ensuremath{\PK}\xspace}
  \def\Kbar  {\kern 0.2em\overline{\kern -0.2em \PK}{}\xspace}

\def\Kz    {\ensuremath{\kaon^0}\xspace}
\def\Kzb   {\ensuremath{\Kbar^0}\xspace}
\def\KzKzb {\ensuremath{\Kz \kern -0.16em \Kzb}\xspace}
\def\Kp    {\ensuremath{\kaon^+}\xspace}
\def\Km    {\ensuremath{\kaon^-}\xspace}

\def\KpKm  {\ensuremath{\Kp \kern -0.16em \Km}\xspace}
\def\KS    {\ensuremath{\kaon^0_{\rm\scriptscriptstyle S}}\xspace}

  \def\Dbar    {\kern 0.2em\overline{\kern -0.2em \PD}{}\xspace}
\def\D       {\ensuremath{\PD}\xspace}

\def\Dz      {\ensuremath{\D^0}\xspace}
\def\Dzb     {\ensuremath{\Dbar^0}\xspace}
\def\DzDzb   {\ensuremath{\Dz {\kern -0.16em \Dzb}}\xspace}
\def\Dp      {\ensuremath{\D^+}\xspace}
\def\Dm      {\ensuremath{\D^-}\xspace}

\def\DpDm    {\ensuremath{\Dp {\kern -0.16em \Dm}}\xspace}
\def\Dstar   {\ensuremath{\D^*}\xspace}

\def\Dstarp  {\ensuremath{\D^{*+}}\xspace}
\def\Dstarm  {\ensuremath{\D^{*-}}\xspace}

\def\B       {\ensuremath{\PB}\xspace}
  \def\Bbar    {\kern 0.18em\overline{\kern -0.18em \PB}{}\xspace}

\def\Bs      {\ensuremath{\B^0_\squark}\xspace}
\def\Bsb     {\ensuremath{\Bbar^0_\squark}\xspace}

  \def\Y#1S{\ensuremath{\PUpsilon{(#1S)}}\xspace}% no space before {...}!

\def\Lbar {\ensuremath{\kern 0.1em\overline{\kern -0.1em\PLambda}}\xspace}

\def\to                 {\ensuremath{\rightarrow}\xspace}

\def\CP                {\ensuremath{C\!P}\xspace}

\def\AT#1     {\ensuremath{A_{\mathrm{T}}^{#1}}\xspace}           % 2

\def\C#1      {\ensuremath{\mathcal{C}_{#1}}\xspace}                       % 9
\def\Cp#1     {\ensuremath{\mathcal{C}_{#1}^{'}}\xspace}                    % 7
\def\Ceff#1   {\ensuremath{\mathcal{C}_{#1}^{\mathrm{(eff)}}}\xspace}        % 9  
\def\Cpeff#1  {\ensuremath{\mathcal{C}_{#1}^{'\mathrm{(eff)}}}\xspace}       % 7
\def\Ope#1    {\ensuremath{\mathcal{O}_{#1}}\xspace}                       % 2
\def\Opep#1   {\ensuremath{\mathcal{O}_{#1}^{'}}\xspace}                    % 7

\newcommand{\tev}{\ensuremath{\mathrm{\,Te\kern -0.1em V}}\xspace}
\newcommand{\gev}{\ensuremath{\mathrm{\,Ge\kern -0.1em V}}\xspace}
\newcommand{\mev}{\ensuremath{\mathrm{\,Me\kern -0.1em V}}\xspace}
\newcommand{\kev}{\ensuremath{\mathrm{\,ke\kern -0.1em V}}\xspace}
\newcommand{\ev}{\ensuremath{\mathrm{\,e\kern -0.1em V}}\xspace}
\newcommand{\gevc}{\ensuremath{{\mathrm{\,Ge\kern -0.1em V\!/}c}}\xspace}
\newcommand{\mevc}{\ensuremath{{\mathrm{\,Me\kern -0.1em V\!/}c}}\xspace}
\newcommand{\gevcc}{\ensuremath{{\mathrm{\,Ge\kern -0.1em V\!/}c^2}}\xspace}
\newcommand{\gevgevcccc}{\ensuremath{{\mathrm{\,Ge\kern -0.1em V^2\!/}c^4}}\xspace}
\newcommand{\mevcc}{\ensuremath{{\mathrm{\,Me\kern -0.1em V\!/}c^2}}\xspace}

\def\invfb   {\ensuremath{\mbox{\,fb}^{-1}}\xspace}

\def\gsim{{~\raise.15em\hbox{$>$}\kern-.85em
          \lower.35em\hbox{$\sim$}~}\xspace}
\def\lsim{{~\raise.15em\hbox{$<$}\kern-.85em
          \lower.35em\hbox{$\sim$}~}\xspace}

\def\tell1  {TELL1\xspace}
\def\ukl1   {UKL1\xspace}

\newcommand{\massmev}{\mevcc}
\newcommand{\pgev}{\gevc}

\newcommand{\pis}{\ensuremath{\pi_{\rm s}}\xspace}
\newcommand{\M}{\ensuremath{M(\Dz\pis^+)}\xspace}
\newcommand{\tos}{\ensuremath{\text{TOS}}\xspace}
\newcommand{\nottos}{\ensuremath{\overline{\text{TOS}}}\xspace}
\newcommand{\Ripred}{\ensuremath{\widetilde{R}_i}}

\newcommand{\chisquaredexpression}{
\begin{equation}\label{eqn:fit}
\chi^2 = \sum_i \left[\left(\frac{ r^+_i-\epsilon_r^+\Ripred^+}{\sigma_i^+}\right)^2+
\left(\frac{r^-_i-\epsilon_r^-\Ripred^-}{\sigma_i^-}\right)^2\right]+\chi^2_\epsilon \, 
+ \, \chi^2_{B} + \, \chi^2_{\rm p} \, .
\end{equation}
}

%% file: title-LHCb-PAPER.tex
% $Id: title-LHCb-PAPER.tex 39841 2013-07-26 10:31:08Z roldeman $
% ===============================================================================
% Purpose: LHCb-PAPER journal paper title page template
% Author: 
% Created on: 2010-09-25
% ===============================================================================

%%%%%%%%%%%%%%%%%%%%%%%%%
%%%%%  TITLE PAGE  %%%%%%
%%%%%%%%%%%%%%%%%%%%%%%%%
\begin{titlepage}
\pagenumbering{roman}

% Header ---------------------------------------------------
\vspace*{-1.5cm}
\centerline{\large EUROPEAN ORGANIZATION FOR NUCLEAR RESEARCH (CERN)}
\vspace*{1.5cm}
\hspace*{-0.5cm}
\begin{tabular*}{\linewidth}{lc@{\extracolsep{\fill}}r}
\vspace*{-1.2cm}\mbox{\!\!\!\includegraphics[width=.12\textwidth]{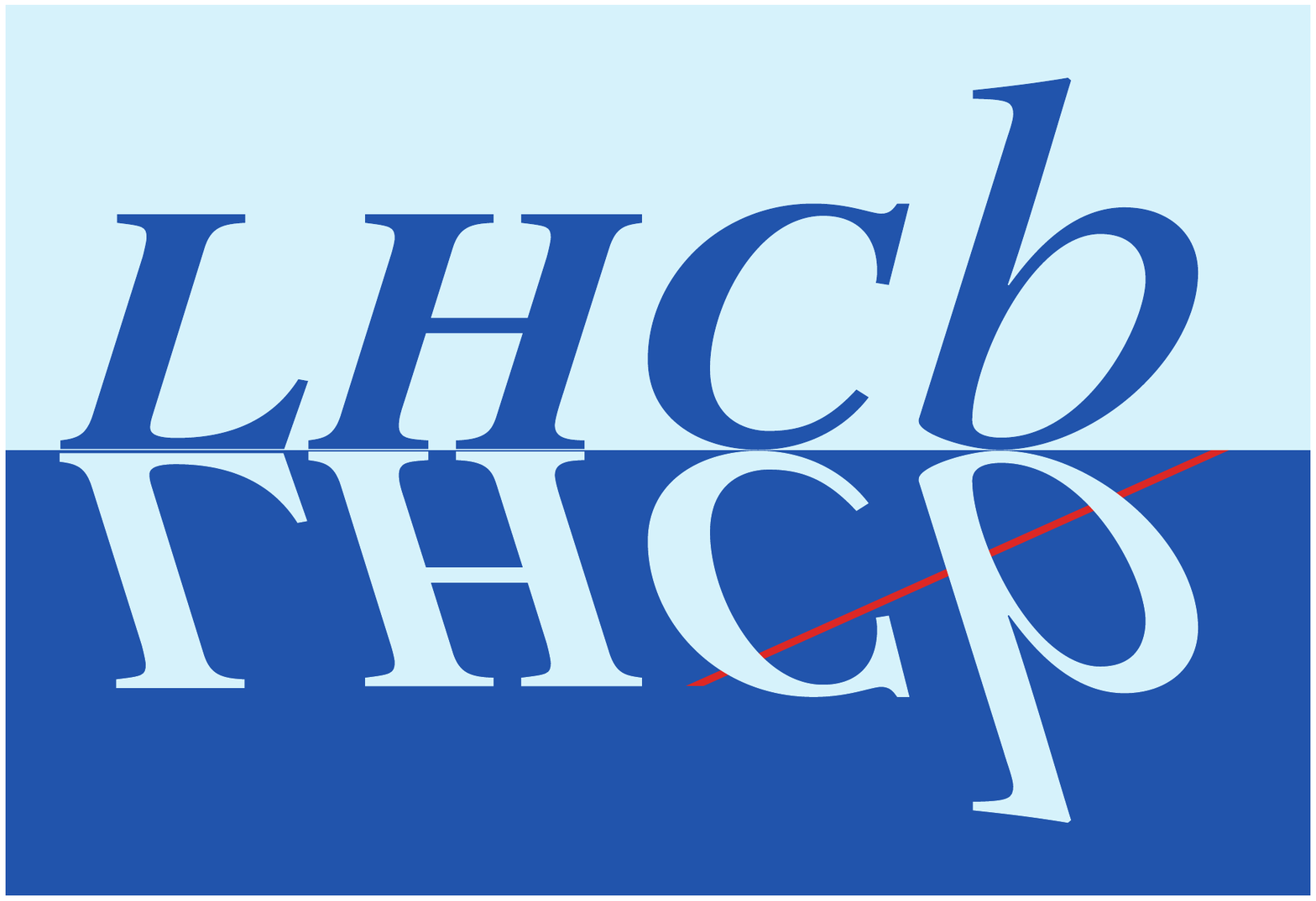}} & & \\
 & & CERN-PH-EP-2013-176 \\  % ID 
 & & LHCb-PAPER-2013-053 \\  % ID 
 & & 18 December 2013 \\ % Date
 & & \\
% not in paper \hline
\end{tabular*}

\vspace*{2.0cm}

% Title --------------------------------------------------
{\bf\boldmath\huge
\begin{center}
Measurement of \Dz--\Dzb mixing parameters and search for \CP violation using $\Dz\to K^+\pi^-$ decays
\end{center}
}

\vspace*{1.0cm}

% Authors -------------------------------------------------
\begin{center}
The LHCb collaboration\footnote{Authors are listed on the following pages.}
\end{center}

\vspace{\fill}

% Abstract -----------------------------------------------
\input{abstract}

\vspace{\fill}

{\footnotesize 
\centerline{\copyright~CERN on behalf of the \lhcb collaboration, license \href{http://creativecommons.org/licenses/by/3.0/}{CC-BY-3.0}.}}
\vspace*{2mm}

\end{titlepage}

%%%%%%%%%%%%%%%%%%%%%%%%%%%%%%%%
%%%%%  EOD OF TITLE PAGE  %%%%%%
%%%%%%%%%%%%%%%%%%%%%%%%%%%%%%%%

%  empty page follows the title page ----
\newpage
\setcounter{page}{2}
\mbox{~}
\newpage

% Author List ----------------------------
\input{LHCb_HD_authorlist_2013-08-07}
\cleardoublepage

%% file: abstract.tex
\begin{abstract}
\noindent 
Measurements of charm mixing parameters from the decay-time-dependent ratio of $\Dz\to K^+\pi^-$ to $\Dz\to K^-\pi^+$ rates and the charge-conjugate ratio are reported. The analysis uses data, corresponding to $3 \invfb$ of integrated luminosity, from proton-proton collisions at 7 and 8 \tev center-of-mass energies recorded by the \lhcb experiment. In the limit of charge-parity (\CP) symmetry, the mixing parameters are determined to be $x'^2=(5.5 \pm 4.9)\times10^{-5}$, $y'= (4.8 \pm 1.0)\times 10^{-3}$, and $R_D=(3.568 \pm 0.066)\times10^{-3}$. Allowing for \CP violation, the mixing parameters are determined separately for \Dz and \Dzb mesons yielding $A_D = (-0.7 \pm 1.9)\%$, for the direct \CP-violating asymmetry, and $0.75 < |q/p|< 1.24$ at the $68.3\%$ confidence level, where $q$ and $p$ are parameters that describe the mass eigenstates of the neutral charm mesons in terms of the flavor eigenstates. This is the most precise determination of these parameters from a single experiment and shows no evidence for \CP violation.
\end{abstract}

%% file: LHCb_HD_authorlist_2013-08-07.tex
%%%%%%%%%%%%%%%%%%%%%%%%%%%%%%%%%%%%%%%%%%
\centerline{\large\bf LHCb collaboration}
\begin{flushleft}
\small
R.~Aaij$^{40}$, 
B.~Adeva$^{36}$, 
M.~Adinolfi$^{45}$, 
C.~Adrover$^{6}$, 
A.~Affolder$^{51}$, 
Z.~Ajaltouni$^{5}$, 
J.~Albrecht$^{9}$, 
F.~Alessio$^{37}$, 
M.~Alexander$^{50}$, 
S.~Ali$^{40}$
% ...

{\footnotesize \it
% ...
$ ^{5}$Clermont Universit\'{e}, Universit\'{e} Blaise Pascal, CNRS/IN2P3, LPC, Clermont-Ferrand, France\\
$ ^{6}$CPPM, Aix-Marseille Universit\'{e}, CNRS/IN2P3, Marseille, France\\
% ...
$ ^{9}$Fakult\"{a}t Physik, Technische Universit\"{a}t Dortmund, Dortmund, Germany\\
% ...
$ ^{36}$Universidad de Santiago de Compostela, Santiago de Compostela, Spain\\
$ ^{37}$European Organization for Nuclear Research (CERN), Geneva, Switzerland\\
% ...
$ ^{40}$Nikhef National Institute for Subatomic Physics, Amsterdam, The Netherlands\\
% ...
$ ^{45}$H.H. Wills Physics Laboratory, University of Bristol, Bristol, United Kingdom\\
% ...
$ ^{50}$School of Physics and Astronomy, University of Glasgow, Glasgow, United Kingdom\\
$ ^{51}$Oliver Lodge Laboratory, University of Liverpool, Liverpool, United Kingdom\\
% ...
}
\end{flushleft}
%%%%%%%%%%%%%%%%%%%%%%%%%%%%%%%%%%%%%%%%%%

%% file: body.tex
%% Introduction and motivation
Mass eigenstates of neutral charm mesons are linear combinations of flavor eigenstates $|D_{1, 2} \rangle = p | \Dz \rangle \pm  q|\Dzb \rangle$, where $p$ and $q$ are complex parameters. This results in \Dz--\Dzb  oscillation. In the limit of charge-parity (\CP) symmetry, the oscillation is characterized by the difference in mass $\Delta m\equiv m_2 - m_1$ and decay width $\Delta\Gamma\equiv\Gamma_2 - \Gamma_1$ between the $D$ mass eigenstates. These differences are usually expressed in terms of the dimensionless mixing parameters $x \equiv \Delta m/\Gamma$ and $y \equiv \Delta\Gamma/2\Gamma$, where $\Gamma$ is the average decay width of neutral \D mesons. If \CP symmetry is violated, the oscillation rates for mesons produced as \Dz and \Dzb can differ, further enriching the phenomenology. Both short- and long-distance components of the amplitude contribute to the time evolution of neutral \D mesons~\cite{Bianco:2003vb,Burdman:2003rs,Artuso:2008vf}. Short-distance amplitudes could include contributions from non-standard-model particles or interactions, possibly enhancing the average oscillation rate or the difference between \Dz and \Dzb meson rates. The study of \CP violation in \Dz oscillation may lead to an improved understanding of possible dynamics beyond the standard model~\cite{Blaylock:1995ay,Petrov:2006nc,Golowich:2007ka,Ciuchini:2007cw}.

% Current experimental status
The first evidence for \Dz--\Dzb oscillation was reported in 2007~\cite{Aubert:2007wf,Staric:2007dt}. By 2009, the hypothesis of no oscillation was excluded with significance in excess of 10 standard deviations~\cite{HFAG} by combining results from different experiments~\cite{Aubert:2007wf,Staric:2007dt,Aaltonen:2007ac,Zhang:2006dp,Abe:2007rd,Aubert:2008zh,Aubert:2009ai,
delAmoSanchez:2010xz,Asner:2012xb}. In 2012, the \lhcb experiment reported the first observation from a single measurement with greater than 5 standard deviation significance~\cite{LHCb-PAPER-2012-038}, which has been recently confirmed by the CDF experiment~\cite{Aaltonen:2013pja}.
 
% Outline and phenomenology details
This Letter reports a search for \CP violation in \Dz--\Dzb mixing by comparing the decay-time-dependent ratio of $\Dz\to K^+\pi^-$ to $\Dz\to K^-\pi^+$ rates with the corresponding ratio for the charge-conjugate processes.  An improved determination of the \CP-averaged charm mixing parameters with respect to our previous measurement~\cite{LHCb-PAPER-2012-038} is also reported. The analysis uses data corresponding to $1.0\invfb$ of integrated luminosity from $\sqrt{s}=7\,\tev$ $pp$ collisions recorded by \lhcb during 2011 and $2.0\invfb$ from $\sqrt{s}=8\,\tev$ collisions recorded during 2012. The neutral \D flavor at production is determined from the charge of the low-momentum pion $\pis^{+}$ in the flavor-conserving strong-interaction decay $\Dstarp\to\Dz\pis^+$.  The inclusion of charge-conjugate processes is implicit unless stated otherwise. The $\Dstarp\to\Dz(\to K^-\pi^+)\pis^+$ process is denoted as right sign (RS), and $\Dstarp\to\Dz(\to K^+\pi^-)\pis^+$ is denoted as wrong sign (WS). The RS decay rate is dominated by a Cabibbo-favored amplitude. The WS rate arises from the interfering amplitudes of the doubly Cabibbo-suppressed $\Dz\to K^+\pi^-$ decay and the Cabibbo-favored $\Dzb\to K^+\pi^-$ decay following \Dz--\Dzb oscillation, each of similar magnitude.  In the limit of $|x|,|y|\ll1$, and assuming negligible \CP violation, the time-dependent ratio $R(t)$ of WS-to-RS decay rates is~\cite{Bianco:2003vb,Burdman:2003rs,Artuso:2008vf,Blaylock:1995ay}
\begin{equation}\label{eq:true-ratio}
R(t) \approx R_D+\sqrt{R_D}\ y'\ \frac{t}{\tau}+\frac{x'^2+y'^2}{4}\left(\frac{t}{\tau}\right)^2,
\end{equation}
where $t$ is the decay time, $\tau$ is the average \Dz lifetime, and $R_D$ is the ratio of suppressed-to-favored decay rates. The  parameters $x'$ and $y'$ depend linearly on the mixing parameters as $x' \equiv x\cos\delta+y\sin\delta$ and $y' \equiv y\cos\delta-x\sin\delta$, where $\delta$ is the strong-phase difference between the suppressed and favored amplitudes $\mathcal{A}(\Dz\to K^+\pi^-)/\mathcal{A}(\Dzb\to K^+\pi^-) = -\sqrt{R_D} e^{-i\delta}$. Allowing for \CP violation, the WS rates $R^+(t)$ and $R^-(t)$ of initially produced \Dz and \Dzb mesons are functions of independent sets of mixing parameters $(R_D^\pm,\, x'^{2\pm},\, y'^\pm)$. A difference between $R^+_D$ and $R^-_D$ arises if the ratio between the magnitudes of suppressed and favored decay amplitudes is not \CP symmetric (direct \CP violation). Violation of \CP symmetry either in mixing $|q/p|\neq1$ or in the interference between mixing and decay amplitudes $\phi\equiv\arg\left[q\mathcal{A}(\Dzb\to K^+\pi^-)/p\mathcal{A}(\Dz\to K^+\pi^-)\right]-\delta\neq0$ are usually referred to as indirect \CP violation and would result in differences between $(x'^{2+},\, y'^+)$ and $(x'^{2-},\, y'^-)$.

% LHCb detector
The \lhcb detector~\cite{Alves:2008zz} is a single-arm forward spectrometer covering the \mbox{pseudorapidity} range $2<\eta <5$, designed for the study of particles containing \bquark or \cquark quarks. Detector components particularly relevant for this analysis are the silicon vertex detector, which provides reconstruction of displaced vertices of \bquark- and \cquark-hadron decays; the tracking system, which measures charged particle momenta with relative uncertainty that varies from $0.4\%$ at $5\,\pgev$ to $0.6\%$ at $100\,\pgev$, corresponding to a typical mass resolution of approximately $8\,\massmev$ for a two-body charm-meson decay; and the ring-imaging Cherenkov detectors, which provide kaon-pion discrimination~\cite{LHCb-DP-2012-003}. The magnet polarity is periodically inverted and approximately equal amounts of data are collected in each configuration to mitigate the effects of detection asymmetries. The online event-selection system (trigger)~\cite{LHCb-DP-2012-004} consists of a first-level hardware stage based on information from the calorimeter and muon systems, followed by a software high-level trigger.

%% Selection and reconstruction
Events with \Dstarp candidates consistent with being produced at the $pp$ collision point (primary vertex) are selected following Ref.~\cite{LHCb-PAPER-2012-038}. In addition, a WS candidate is discarded if resulting from a $D^0$ candidate that, associated with another pion, also forms a RS candidate with \M within $3\,\massmev$ of the known \Dstarp mass. This removes about 15\% of the WS background with negligible signal loss. The two-body $D^0\pis^+$ mass \M is computed using the known \Dz and $\pi^+$ masses~\cite{PDG2012} and their reconstructed momenta~\cite{LHCb-PAPER-2012-038}. In Ref.~\cite{LHCb-PAPER-2012-038}, we used events selected by the hardware trigger based on hadron calorimeter transverse-energy depositions that were geometrically matched with signal final-state tracks. In the present analysis, we distinguish two trigger categories. One category consists of events that meet the above trigger requirement (triggered-on-signal, \tos). The other comprises events with candidates failing the track-calorimeter matching and events selected based on muon hardware triggers decisions (\nottos). The two subsamples contribute approximately equal signal yields with similar purities. However, they require separate treatment due to their differing kinematic distributions and trigger-induced biases.

%%Yield determination
\begin{figure}[h!]
\centering
\includegraphics[width=0.4\textwidth]{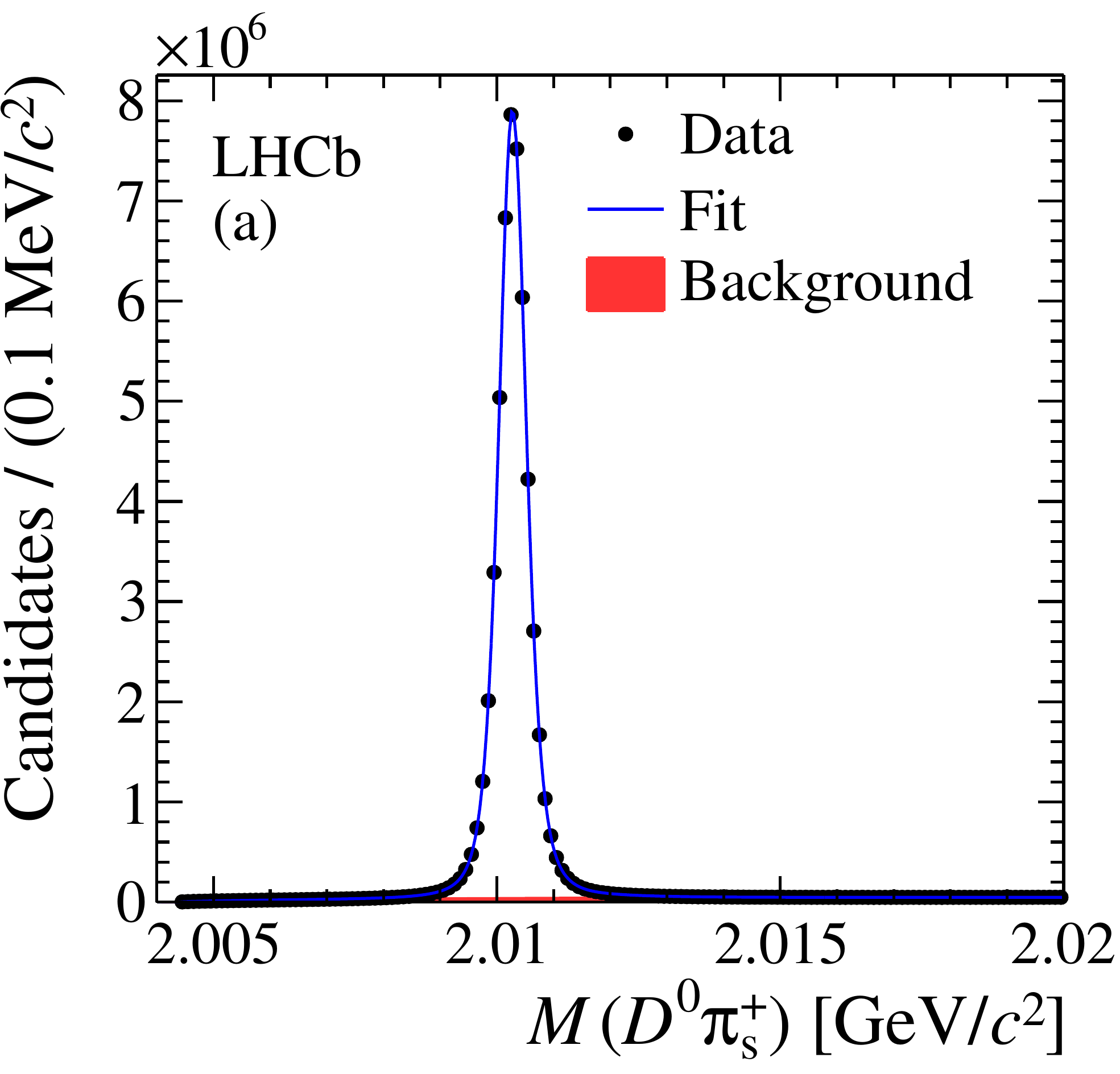}
\includegraphics[width=0.4\textwidth]{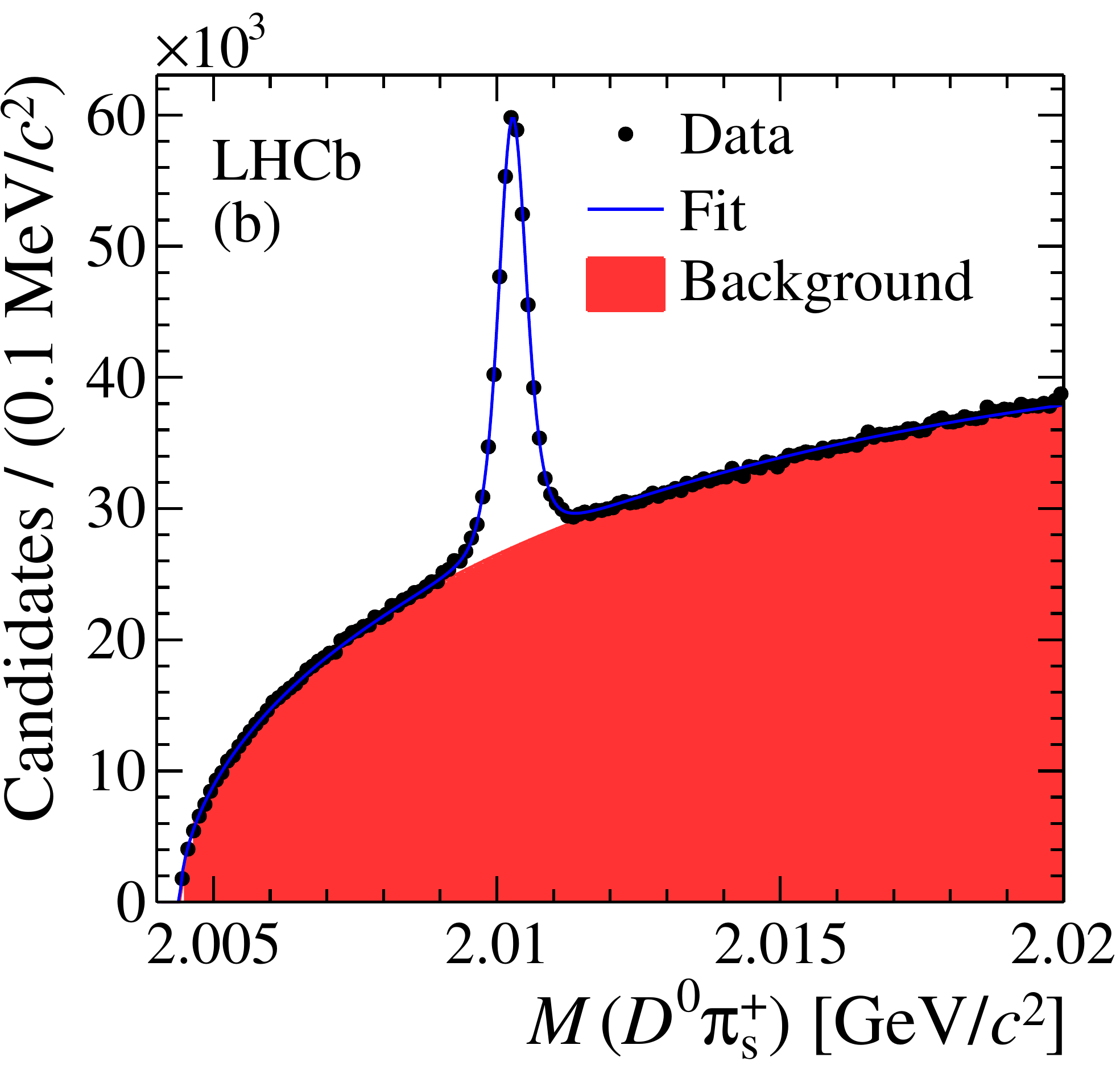}
\caption{\small Distribution of \M  for selected (a) right-sign $\Dz\to K^-\pi^+$ and (b) wrong-sign $\Dz\to K^+\pi^-$ candidates.\label{fig:mass}}
\end{figure}

\begin{figure}[t!]
\centering
\includegraphics[width=0.5\textwidth]{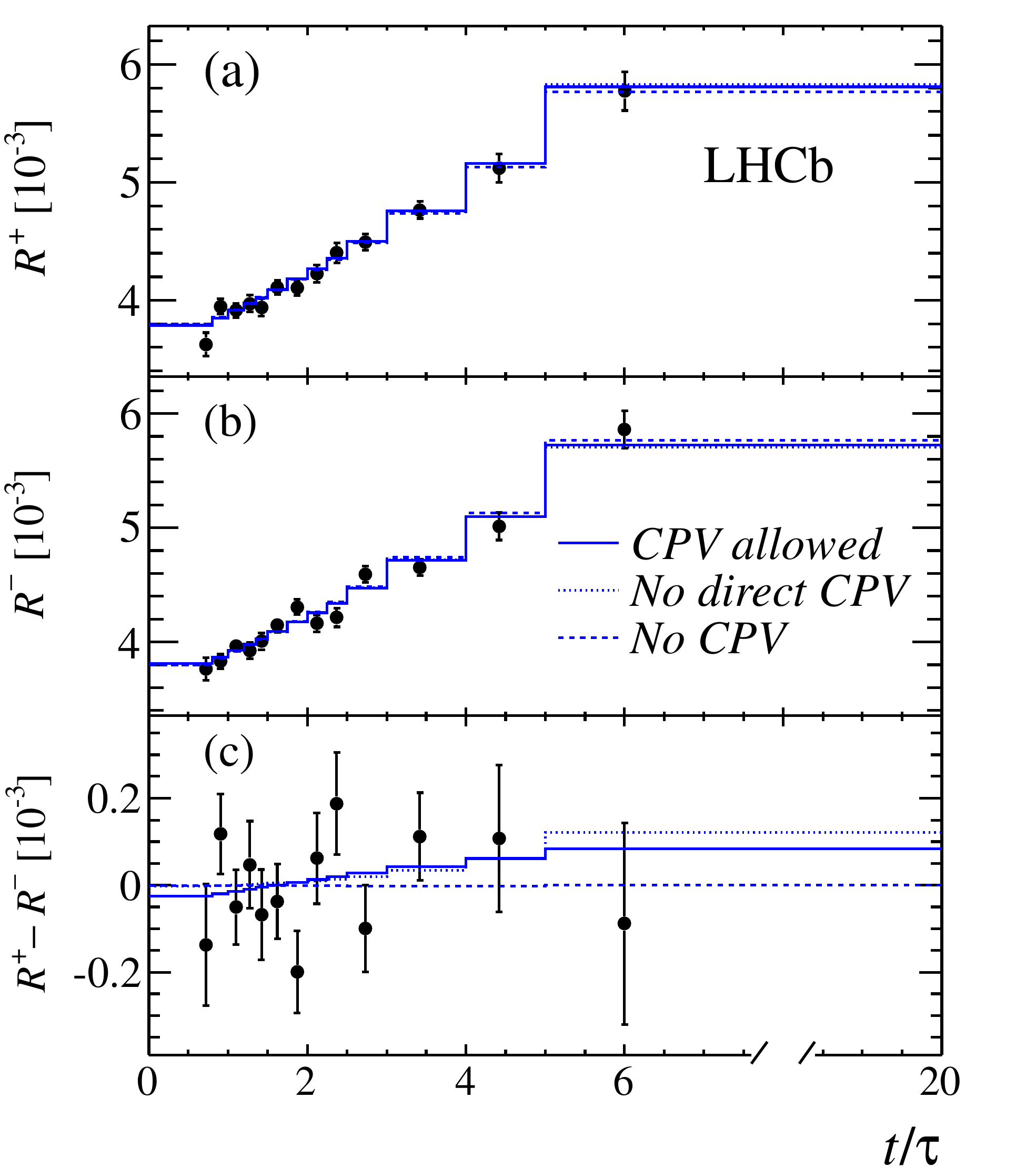}
\caption{\small Efficiency-corrected ratios of WS-to-RS yields for (a) \Dstarp decays, (b) \Dstarm decays, and (c) their differences as functions of decay time in units of \Dz lifetime. Projections of fits allowing for (dashed line) no \CP violation, (dotted line) no direct \CP violation, and (solid line) full \CP violation are overlaid. The abscissa of the data points corresponds to the average decay time over the bin; the error bars indicate the statistical uncertainties.\label{fig:finalResults}}
\end{figure}

The RS and WS signal yields are determined by fitting the \M distribution of \Dz candidates with reconstructed mass within $24\,\massmev$ of the known value. The time-integrated \M distributions are shown in Fig.~\ref{fig:mass}. The smooth background is dominated by favored $\Dzb\to K^+\pi^-$ decays associated with random $\pis^+$ candidates. The sample contains $1.15 \times 10^5$ ($1.14 \times 10^5$) signal WS \Dz (\Dzb) decays and approximately 230 times more RS decays. Yield differences between \Dz and \Dzb decays are dominated by differences in charm-anticharm production rates and reconstruction efficiencies. Each sample is divided into 13 subsamples according to the candidate's decay time, and signal yields are determined for each using shape parametrizations determined from simulation and tuned to data~\cite{LHCb-PAPER-2012-038}. 
We assume that for a given \Dstar meson flavor, the signal shapes are common to WS and RS decays, while the descriptions of the background can differ. The decay-time-dependent WS-to-RS yield ratios $R^+$ and $R^-$ observed in the \Dz and \Dzb samples, respectively, and their difference are shown in Fig.~\ref{fig:finalResults}. These are corrected for the relative efficiencies for reconstructing $K^- \pi^+$ and $K^+\pi^-$ final states.

%% chi^2 epression
The mixing parameters are determined by minimizing a $\chi^2$ variable that includes terms for the difference between the observed and predicted ratios and for systematic deviations of parameters
\chisquaredexpression
The measured WS-to-RS yield ratio and its statistical uncertainty in the decay-time bin $i$ are denoted by $r_i^{\pm}$ and $\sigma_i^{\pm}$, respectively. The predicted value for the WS-to-RS yield ratio $\Ripred^{\pm}$ corresponds to the time integral over bin $i$ of Eq.~\eqref{eq:true-ratio} including bin-specific corrections. These account for small biases due to the decay-time evolution of the approximately $3\%$ fraction of signal candidates originating from \bquark-hadron decays ($\Delta_B$) and of the about $0.5\%$ component of peaking background from RS decays in which both final-state particles are misidentified ($\Delta_{\rm p}$)~\cite{LHCb-PAPER-2012-038}. The relative efficiency $\epsilon_r^\pm$ accounts for instrumental asymmetries in the $K\pi$ reconstruction efficiencies, mainly caused by $K^-$ mesons having a larger interaction cross section with matter than $K^+$ mesons. These asymmetries are measured in data to be in the range $0.8$--$1.2\%$ with $0.2\%$ precision and to be independent of decay time. They are derived from the efficiency ratio $\epsilon_r^+=1/\epsilon_r^-=\epsilon(K^+\pi^-)/\epsilon(K^-\pi^+)$, obtained from the product of $D^- \to K^+\pi^-\pi^-$ and $D^+ \to \KS (\to \pi^+\pi^-)\pi^+$ event yields divided by the product of the corresponding charge-conjugate decay yields. No \CP violation is expected or experimentally observed~\cite{PDG2012} in these decays. Asymmetries due to \CP violation in neutral kaons and their interaction cross-sections with matter are negligible. The $1\%$ asymmetry between \Dp  and \Dm production rates~\cite{LHCb-PAPER-2012-026} cancels in this ratio, provided that the kinematic distributions are consistent across samples. We weight the $D^- \to K^+ \pi^- \pi^-$ events so that their kinematic distributions match those in the $D^+ \to \KS \pi^+$ sample. Similarly, these samples are weighted as functions of $K\pi$ momentum to match the RS momentum spectra. The parameters associated with $\Delta_B$, $\Delta_{\rm p}$, and $\epsilon_r$ are determined separately for \tos and \nottos subsets and vary independently in the fit within their Gaussian constraints $\chi^2_B$, $\chi^2_{\rm p}$, and $\chi^2_\epsilon$~\cite{LHCb-PAPER-2012-038}.

% Blinding and tests
To avoid experimenters' bias in the \CP violation parameters, the measurement technique is finalized by adding arbitrary offsets to the WS-to-RS yield ratios for the \Dz and \Dzb samples, designed to mimic the effect of different mixing parameters in the two samples. To rule out global systematic uncertainties not accounted for in Eq.~\eqref{eqn:fit}, the data are first integrated over the whole decay-time spectrum and subsequently divided into statistically independent subsets according to criteria likely to reveal biases from specific instrumental effects. These include the number of primary vertices in the events, the $K$ laboratory momentum, the \pis impact parameter $\chi^2$ with respect to the primary vertex, the \Dz impact parameter $\chi^2$ with respect to the primary vertex, the magnetic field orientation, and the hardware trigger category. The variations of the time-integrated charge asymmetry in WS-to-RS yield ratios are consistent with statistical fluctuations. Then, we investigate decay-time-dependent biases by dividing the time-binned sample according to the magnet polarity and the number of primary vertices per event. In the \tos sample, differences of WS-to-RS yield ratios as functions of decay time for opposite magnet polarities yield $\chi^2$ values of 12, 17, and 14 (for 12  degrees of freedom), for events with one, two, and more than two primary vertices, respectively. The corresponding $\chi^2$ values in the \nottos sample, 9, 11, and 8, suggest a systematically better consistency. Hence, the statistical uncertainty of each of the WS-to-RS ratios in the \tos samples is increased by a factor of $\sqrt{17/12}$, following Ref.~\cite{PDG2012}. These scaled uncertainties are used in all subsequent fits. Independent analyses of the 2011 and 2012 data yield consistent results. The ratio between RS \Dz to \Dzb decay rates is independent of decay time with a $62\%$ $p$ value and a standard deviation of $0.16\%$, showing no evidence of correlations between particle identification or reconstruction efficiency and decay time.

\begin{table}
\centering
\caption{\small Results of fits to the data for different hypotheses on the \CP symmetry. The reported uncertainties are statistical and systematic, respectively; ndf indicates the number of degrees of freedom. See App.~\ref{app:supp} for fits results including correlation coefficients.\label{tab:finalResults}}
\begin{tabular}{l@{ [}lr@{\,$\pm$\,}c@{\,$\pm$\,}l}
\hline\hline
\multicolumn{2}{l}{Parameter} & \multicolumn{3}{c}{Value}\\
\hline
\multicolumn{5}{c}{Direct and indirect \CP violation}\\
$R_D^+$ &$10^{-3}$] & $3.545$ & $0.082$ & $0.048$ \\
$y'^+$     &$10^{-3}$] & $5.1$ & $1.2$ & $0.7$ \\
$x'^{2+}$ &$10^{-5}$] & $4.9$ & $6.0$ & $3.6$ \\
$R_D^-$  &$10^{-3}$] & $3.591$ & $0.081$ & $0.048$ \\
$y'^-$      &$10^{-3}$] & $4.5$ & $1.2$ & $0.7$ \\
$x'^{2-}$  &$10^{-5}$] & $6.0$ & $5.8$ & $3.6$ \\
\multicolumn{2}{l}{$\chi^2/\text{ndf}$}  & \multicolumn{3}{c}{$85.9/98$} \\
\multicolumn{5}{c}{}\\
\multicolumn{5}{c}{No direct \CP violation}\\
$R_D$      &$10^{-3}$] & $3.568$ & $0.058$ & $0.033$ \\
$y'^+$     &$10^{-3}$] & $4.8$ & $0.9$ & $0.6$ \\
$x'^{2+}$ &$10^{-5}$] & $6.4$ & $4.7$ & $3.0$ \\
$y'^-$      &$10^{-3}$] & $4.8$ & $0.9$ & $0.6$ \\
$x'^{2-}$  &$10^{-5}$] & $4.6$ & $4.6$ & $3.0$ \\
\multicolumn{2}{l}{$\chi^2/\text{ndf}$}  & \multicolumn{3}{c}{$86.0/99$} \\
\multicolumn{5}{c}{}\\
\multicolumn{5}{c}{No \CP violation}\\
$R_D$ &$10^{-3}$] & $3.568$ & $0.058$ & $0.033$ \\
$y'$     &$10^{-3}$] & $4.8$ & $0.8$ & $0.5$ \\
$x'^2$ &$10^{-5}$] & $5.5$ & $4.2$ & $2.6$\\
\multicolumn{2}{l}{$\chi^2/\text{ndf}$} & \multicolumn{3}{c}{$86.4/101$} \\
\hline\hline
\end{tabular}
\end{table}

\begin{figure*}[t]
\centering
\includegraphics[width=\textwidth]{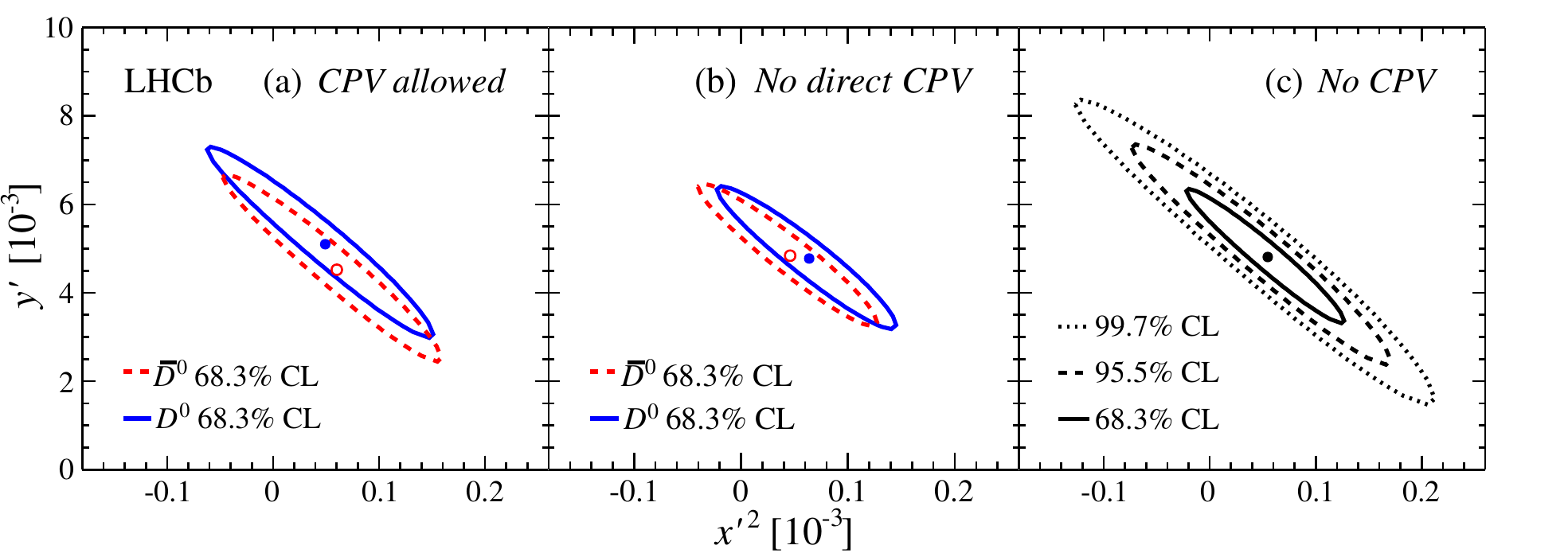}
\caption{\small Two-dimensional confidence regions in the $(x'^2,y')$ plane obtained (a) without any restriction on \CP violation, (b) assuming 
no direct \CP violation,  and (c) assuming \CP conservation. The dashed (solid) curves in (a) and (b) indicate the contours of the mixing parameters associated with \Dzb (\Dz) decays. The best-fit value for \Dzb (\Dz) decays is shown with an open (filled) point. The solid, dashed, and dotted curves in (c) indicate the contours of \CP-averaged mixing parameters at 68.3\%, 95.5\%, and 99.7\% confidence level (CL), respectively. The best-fit value is shown with a point.
\label{fig:contours}}
\end{figure*}

Three fits are performed to the data shown in Fig.~\ref{fig:finalResults}. The first allows direct and indirect \CP violation; the second allows only indirect \CP violation by constraining $R_D^\pm$ to a common value; and the third is a \CP-conserving fit that constrains all mixing parameters to be the same in the \Dz and \Dzb samples. The fit results and their projections are shown in Table~\ref{tab:finalResults} and Fig.~\ref{fig:finalResults}, respectively. Figure~\ref{fig:contours} shows the central values and confidence regions in the $(x'^2,\, y')$ plane. For each fit, 104 WS-to-RS ratio data points are used, corresponding to 13 ranges of decay time, distinguishing \Dstarp from \Dstarm decays, \tos from \nottos decays, and 2011 data from 2012 data. The consistency with the hypothesis of \CP symmetry is determined from the change in $\chi^2$ between the fit without and with \CP violation, taking into account the difference in number of degrees of freedom. The resulting $p$ value, for the fit with direct and indirect (indirect only) \CP violation allowed, is $91\%$ ($81\%$), showing that the data are compatible with \CP symmetry.

The uncertainties incorporate both statistical and systematic contributions, since all relevant systematic effects depend on the true values of the mixing parameters, and are thus incorporated into the fit $\chi^2$. These include the uncertainty in the fraction of charm mesons from $b$-hadron decays, and their bias on the observed decay time; the uncertainty in the fraction of peaking background; and the uncertainty in the determination of the instrumental asymmetry. The statistical uncertainty is determined in a separate fit and used to calculate the systematic component by subtraction in quadrature.

Direct \CP violation would produce a nonzero intercept at $t=0$ in the efficiency-corrected difference of WS-to-RS yield ratios between \Dz and \Dzb mesons shown in Fig.~\ref{fig:finalResults}\,(c). It is parametrized by the asymmetry measured in the first fit $A_D \equiv { (R_D^+ - R_D^-) / (R_D^+ + R_D^-) } =  (-0.7\pm1.9)\%$. Indirect \CP violation results in a time dependence of the efficiency-corrected difference of yield ratios. The slope observed in Fig.~\ref{fig:finalResults}\,(c) is about 5\% of the individual slopes of Figs.~\ref{fig:finalResults}\,(a) and (b) and is consistent with zero. From the results of the fit allowing for direct and indirect \CP violation, a likelihood for $|q/p|$ is constructed using the relations $x'^\pm = |q/p|^{\pm 1}(x'\cos\phi \pm y'\sin\phi)$ and $y'^\pm = |q/p|^{\pm 1}(y'\cos\phi \mp x'\sin\phi)$. Confidence intervals are derived with a likelihood-ratio ordering and assuming that the correlations are independent of the true values of the mixing parameters. The  magnitude of $q/p$ is determined to be $0.75 < |q/p|< 1.24$ and $0.67 < |q/p|< 1.52$ at the $68.3\%$ and $95.5\%$ confidence levels, respectively. Significantly more stringent  bounds on $|q/p|$ and additional information on $\phi$ are available by combining the present results with other measurements~\cite{HFAG}, in particular when also using theoretical constraints, such as the relationship $\tan\phi = x ( 1 - |q/p|^2) / y ( 1+ |q/p|^2)$~\cite{Grossman:2009mn, Kagan:2009gb}, which applies in the limit that direct \CP violation is negligible.

In summary, \Dz--\Dzb oscillation is studied using $\Dstarp\to \Dz (\to K^+\pi^-) \pi^+$ decays reconstructed in the full sample of $pp$ collisions, corresponding to $3\invfb$ of integrated luminosity collected by the \lhcb experiment in 2011 and 2012. Assuming \CP conservation, the mixing parameters are measured to be $x'^2=(5.5 \pm 4.9)\times10^{-5}$, $y'= (4.8 \pm 1.0)\times 10^{-3}$, and $R_D=(3.568 \pm 0.066)\times10^{-3}$. The observed parameters are consistent with, $2.5$ times more precise than, and supersede the results based on a subset of the present data~\cite{LHCb-PAPER-2012-038}. Studying \Dz and \Dzb decays separately shows no evidence for \CP violation and provides the most stringent bounds on the parameters $A_D$ and $|q/p|$ from a single experiment.

%% file: acknowledgements.tex
\section*{Acknowledgements}
\noindent We express our gratitude to our colleagues in the CERN
accelerator departments for the excellent performance of the LHC. We
thank the technical and administrative staff at the LHCb
institutes. We acknowledge support from CERN and from the national
agencies: CAPES, CNPq, FAPERJ and FINEP (Brazil); NSFC (China);
CNRS/IN2P3 and Region Auvergne (France); BMBF, DFG, HGF and MPG
(Germany); SFI (Ireland); INFN (Italy); FOM and NWO (The Netherlands);
SCSR (Poland); MEN/IFA (Romania); MinES, Rosatom, RFBR and NRC
``Kurchatov Institute'' (Russia); MinECo, XuntaGal and GENCAT (Spain);
SNSF and SER (Switzerland); NAS Ukraine (Ukraine); STFC (United
Kingdom); NSF (USA). We also acknowledge the support received from the
ERC under FP7. The Tier1 computing centres are supported by IN2P3
(France), KIT and BMBF (Germany), INFN (Italy), NWO and SURF (The
Netherlands), PIC (Spain), GridPP (United Kingdom). We are thankful
for the computing resources put at our disposal by Yandex LLC
(Russia), as well as to the communities behind the multiple open
source software packages that we depend on.

%% file: appendix.tex
\newpage
\appendix
\section{Supplemental material\label{app:supp}}

\begin{table}[hb!!]
\centering
\caption{\small Detailed fit results. Reported uncertainties and correlation coefficients include both statistical and systematic sources.}
\begin{tabular}{l@{ [}lccccccc}
\hline\hline
\multicolumn{9}{c}{Direct and indirect \CP violation}\\
\multicolumn{3}{c}{Results} & \multicolumn{6}{c}{Correlations} \\
\multicolumn{2}{c}{Parameter} & Fit value & $R_D^+$ & $y'^+$ & $x'^{2+}$ & $R_D^-$ & $y'^{-}$ & $x'^{2-}$ \\
\hline
$R_D^+$ &$10^{-3}$] & $3.545\pm0.095$ & $1.000$ & $-0.942$ & $\phantom{-}0.862$ & $-0.016$ & $-0.007$ & $\phantom{-}0.006$\\
$y'^+$     &$10^{-3}$] & $5.1\pm1.4$            &           &  $\phantom{-}1.000$ & $-0.968$ & $-0.007$ & $\phantom{-}0.007$ & $-0.007$\\
$x'^{2+}$&$10^{-5}$] & $4.9\pm7.0$            &           &             & $\phantom{-}1.000$ & $\phantom{-}0.005$ & $-0.007$ & $\phantom{-}0.008$\\
$R_D^-$  &$10^{-3}$] & $3.591\pm0.094$ &          &              &            &  $\phantom{-}1.000$ & $-0.941$ & $\phantom{-}0.858$\\
$y'^-$      &$10^{-3}$] & $4.5\pm1.4$            &           &              &           &             & $\phantom{-}1.000$ & $-0.966$\\
$x'^{2-}$ &$10^{-5}$] & $6.0\pm6.8$ &          &           &             &             &             &  $\phantom{-}1.000$ \\
\hline\hline
\end{tabular}
\vskip30pt
\begin{tabular}{l@{ [}lcccccc}
\hline\hline
\multicolumn{8}{c}{No direct \CP violation}\\
\multicolumn{3}{c}{Results} & \multicolumn{5}{c}{Correlations} \\
\multicolumn{2}{c}{Parameter} & Fit value & $R_D$ & $y'^+$ &  $x'^{2+}$ & $y'^-$ & $x'^{2-}$ \\
\hline
$R_D$     & $10^{-3}$] & $3.568\pm0.066$ &  $1.000$ & $-0.894$  &  $\phantom{-}0.770$  & $-0.895$ &  $\phantom{-}0.772$ \\
$y'^+$    & $10^{-3}$] & $4.8\pm1.1$       &                &  $\phantom{-}1.000$   &  $-0.949$ & $\phantom{-}0.765$ & $-0.662$\\
$x'^{2+}$ & $10^{-5}$] & $6.4\pm5.5$             &                &                  &   $\phantom{-}1.000$ & $-0.662$ & $\phantom{-}0.574$ \\
$y'^-$      & $10^{-3}$] & $4.8\pm1.1$       &                &                  &                 &  $\phantom{-}1.000$ & $-0.950$ \\
$x'^{2-}$ & $10^{-5}$] & $4.6\pm5.5$            &                 &                  &                 &                &  $\phantom{-}1.000$ \\
\hline\hline
\end{tabular}
\vskip30pt
\begin{tabular}{l@{ [}lcccc}
\hline\hline
\multicolumn{6}{c}{No \CP violation}\\
\multicolumn{3}{c}{Results} & \multicolumn{3}{c}{Correlations} \\
\multicolumn{2}{c}{Parameter} & Fit value & $R_D$ & $y'$ & $x'^2$ \\
\hline
$R_D$  & $10^{-3}$] & $3.568\pm0.066$ & $1.000$ & $-0.953$ & $\phantom{-}0.869$ \\
$y'$     & $10^{-3}$] & $4.8\pm1.0$   &             &  $\phantom{-}1.000$ & $-0.967$ \\
$x'^2$ & $10^{-5}$] & $5.5\pm4.9$ &             &            &   $\phantom{-}1.000$ \\
\hline\hline
\end{tabular}
\end{table}

%% file: main.bbl
\begin{mcitethebibliography}{10}
\mciteSetBstSublistMode{n}
\mciteSetBstMaxWidthForm{subitem}{\alph{mcitesubitemcount})}
\mciteSetBstSublistLabelBeginEnd{\mcitemaxwidthsubitemform\space}
{\relax}{\relax}

\bibitem{Bianco:2003vb}
S.~Bianco, F.~Fabbri, D.~Benson, and I.~Bigi,
  \ifthenelse{\boolean{articletitles}}{{\it {A Cicerone for the physics of
  charm}}, }{}\href{http://dx.doi.org/10.1393/ncr/i2003-10003-1}{Riv.\ Nuovo
  Cim.\  {\bf 26N7} (2003) 1}, \href{http://arxiv.org/abs/hep-ex/0309021}{{\tt
  arXiv:hep-ex/0309021}}\relax
\mciteBstWouldAddEndPuncttrue
\mciteSetBstMidEndSepPunct{\mcitedefaultmidpunct}
{\mcitedefaultendpunct}{\mcitedefaultseppunct}\relax
\EndOfBibitem
\bibitem{Burdman:2003rs}
G.~Burdman and I.~Shipsey, \ifthenelse{\boolean{articletitles}}{{\it {\Dz--\Dzb
  mixing and rare charm decays}},
  }{}\href{http://dx.doi.org/10.1146/annurev.nucl.53.041002.110348}{Ann.\ Rev.\
  Nucl.\ Part.\ Sci.\  {\bf 53} (2003) 431},
  \href{http://arxiv.org/abs/hep-ph/0310076}{{\tt arXiv:hep-ph/0310076}}\relax
\mciteBstWouldAddEndPuncttrue
\mciteSetBstMidEndSepPunct{\mcitedefaultmidpunct}
{\mcitedefaultendpunct}{\mcitedefaultseppunct}\relax
\EndOfBibitem
\bibitem{Artuso:2008vf}
M.~Artuso, B.~Meadows, and A.~A. Petrov,
  \ifthenelse{\boolean{articletitles}}{{\it {Charm meson decays}},
  }{}\href{http://dx.doi.org/10.1146/annurev.nucl.58.110707.171131}{Ann.\ Rev.\
  Nucl.\ Part.\ Sci.\  {\bf 58} (2008) 249},
  \href{http://arxiv.org/abs/0802.2934}{{\tt arXiv:0802.2934}}\relax
\mciteBstWouldAddEndPuncttrue
\mciteSetBstMidEndSepPunct{\mcitedefaultmidpunct}
{\mcitedefaultendpunct}{\mcitedefaultseppunct}\relax
\EndOfBibitem
\bibitem{Blaylock:1995ay}
G.~Blaylock, A.~Seiden, and Y.~Nir, \ifthenelse{\boolean{articletitles}}{{\it
  {The role of CP violation in \Dz--\Dzb mixing}},
  }{}\href{http://dx.doi.org/10.1016/0370-2693(95)00787-L}{Phys.\ Lett.\  {\bf
  B 355} (1995) 555}, \href{http://arxiv.org/abs/hep-ph/9504306}{{\tt
  arXiv:hep-ph/9504306}}\relax
\mciteBstWouldAddEndPuncttrue
\mciteSetBstMidEndSepPunct{\mcitedefaultmidpunct}
{\mcitedefaultendpunct}{\mcitedefaultseppunct}\relax
\EndOfBibitem
\bibitem{Petrov:2006nc}
A.~A. Petrov, \ifthenelse{\boolean{articletitles}}{{\it {Charm mixing in the
  standard model and beyond}},
  }{}\href{http://dx.doi.org/10.1142/S0217751X06034902}{Int.\ J.\ Mod.\ Phys.\
  {\bf A 21} (2006) 5686}, \href{http://arxiv.org/abs/hep-ph/0611361}{{\tt
  arXiv:hep-ph/0611361}}\relax
\mciteBstWouldAddEndPuncttrue
\mciteSetBstMidEndSepPunct{\mcitedefaultmidpunct}
{\mcitedefaultendpunct}{\mcitedefaultseppunct}\relax
\EndOfBibitem
\bibitem{Golowich:2007ka}
E.~Golowich, J.~A. Hewett, S.~Pakvasa, and A.~A. Petrov,
  \ifthenelse{\boolean{articletitles}}{{\it {Implications of \Dz--\Dzb mixing
  for new physics}},
  }{}\href{http://dx.doi.org/10.1103/PhysRevD.76.095009}{Phys.\ Rev.\  {\bf D
  76} (2007) 095009}, \href{http://arxiv.org/abs/0705.3650}{{\tt
  arXiv:0705.3650}}\relax
\mciteBstWouldAddEndPuncttrue
\mciteSetBstMidEndSepPunct{\mcitedefaultmidpunct}
{\mcitedefaultendpunct}{\mcitedefaultseppunct}\relax
\EndOfBibitem
\bibitem{Ciuchini:2007cw}
M.~Ciuchini {\em et~al.}, \ifthenelse{\boolean{articletitles}}{{\it {\Dz--\Dzb
  mixing and new physics: general considerations and constraints on the MSSM}},
  }{}\href{http://dx.doi.org/10.1016/j.physletb.2007.08.055}{Phys.\ Lett.\
  {\bf B 655} (2007) 162}, \href{http://arxiv.org/abs/hep-ph/0703204}{{\tt
  arXiv:hep-ph/0703204}}\relax
\mciteBstWouldAddEndPuncttrue
\mciteSetBstMidEndSepPunct{\mcitedefaultmidpunct}
{\mcitedefaultendpunct}{\mcitedefaultseppunct}\relax
\EndOfBibitem
\bibitem{Aubert:2007wf}
\babar collaboration, B.~Aubert {\em et~al.},
  \ifthenelse{\boolean{articletitles}}{{\it {Evidence for \Dz--\Dzb mixing}},
  }{}\href{http://dx.doi.org/10.1103/PhysRevLett.98.211802}{Phys.\ Rev.\ Lett.\
   {\bf 98} (2007) 211802}, \href{http://arxiv.org/abs/hep-ex/0703020}{{\tt
  arXiv:hep-ex/0703020}}\relax
\mciteBstWouldAddEndPuncttrue
\mciteSetBstMidEndSepPunct{\mcitedefaultmidpunct}
{\mcitedefaultendpunct}{\mcitedefaultseppunct}\relax
\EndOfBibitem
\bibitem{Staric:2007dt}
Belle collaboration, M.~Staric {\em et~al.},
  \ifthenelse{\boolean{articletitles}}{{\it {Evidence for \Dz--\Dzb mixing}},
  }{}\href{http://dx.doi.org/10.1103/PhysRevLett.98.211803}{Phys.\ Rev.\ Lett.\
   {\bf 98} (2007) 211803}, \href{http://arxiv.org/abs/hep-ex/0703036}{{\tt
  arXiv:hep-ex/0703036}}\relax
\mciteBstWouldAddEndPuncttrue
\mciteSetBstMidEndSepPunct{\mcitedefaultmidpunct}
{\mcitedefaultendpunct}{\mcitedefaultseppunct}\relax
\EndOfBibitem
\bibitem{HFAG}
Heavy Flavor Averaging Group, Y.~Amhis {\em et~al.},
  \ifthenelse{\boolean{articletitles}}{{\it {Averages of $b$-hadron,
  $c$-hadron, and $\tau$-lepton properties as of early 2012}},
  }{}\href{http://arxiv.org/abs/1207.1158}{{\tt arXiv:1207.1158}}, {updated
  results and plots available at
  \href{http://www.slac.stanford.edu/xorg/hfag/}{{\tt
  http://www.slac.stanford.edu/xorg/hfag/}}}\relax
\mciteBstWouldAddEndPuncttrue
\mciteSetBstMidEndSepPunct{\mcitedefaultmidpunct}
{\mcitedefaultendpunct}{\mcitedefaultseppunct}\relax
\EndOfBibitem
\bibitem{Aaltonen:2007ac}
CDF collaboration, T.~Aaltonen {\em et~al.},
  \ifthenelse{\boolean{articletitles}}{{\it {Evidence for \Dz--\Dzb mixing
  using the CDF II detector}},
  }{}\href{http://dx.doi.org/10.1103/PhysRevLett.100.121802}{Phys.\ Rev.\
  Lett.\  {\bf 100} (2008) 121802}, \href{http://arxiv.org/abs/0712.1567}{{\tt
  arXiv:0712.1567}}\relax
\mciteBstWouldAddEndPuncttrue
\mciteSetBstMidEndSepPunct{\mcitedefaultmidpunct}
{\mcitedefaultendpunct}{\mcitedefaultseppunct}\relax
\EndOfBibitem
\bibitem{Zhang:2006dp}
Belle collaboration, L.~Zhang {\em et~al.},
  \ifthenelse{\boolean{articletitles}}{{\it {Improved constraints on \Dz--\Dzb
  mixing in $\Dz\to K^+\pi^-$ decays at Belle}},
  }{}\href{http://dx.doi.org/10.1103/PhysRevLett.96.151801}{Phys.\ Rev.\ Lett.\
   {\bf 96} (2006) 151801}, \href{http://arxiv.org/abs/hep-ex/0601029}{{\tt
  arXiv:hep-ex/0601029}}\relax
\mciteBstWouldAddEndPuncttrue
\mciteSetBstMidEndSepPunct{\mcitedefaultmidpunct}
{\mcitedefaultendpunct}{\mcitedefaultseppunct}\relax
\EndOfBibitem
\bibitem{Abe:2007rd}
Belle collaboration, L.~Zhang {\em et~al.},
  \ifthenelse{\boolean{articletitles}}{{\it {Measurement of \Dz--\Dzb mixing
  parameters in $\Dz\to\KS\pi^+\pi^-$ decays}},
  }{}\href{http://dx.doi.org/10.1103/PhysRevLett.99.131803}{Phys.\ Rev.\ Lett.\
   {\bf 99} (2007) 131803}, \href{http://arxiv.org/abs/0704.1000}{{\tt
  arXiv:0704.1000}}\relax
\mciteBstWouldAddEndPuncttrue
\mciteSetBstMidEndSepPunct{\mcitedefaultmidpunct}
{\mcitedefaultendpunct}{\mcitedefaultseppunct}\relax
\EndOfBibitem
\bibitem{Aubert:2008zh}
\babar collaboration, B.~Aubert {\em et~al.},
  \ifthenelse{\boolean{articletitles}}{{\it {Measurement of \Dz--\Dzb mixing
  from a time-dependent amplitude analysis of $\Dz\to K^{+}\pi^{-} \pi^0$
  decays}}, }{}\href{http://dx.doi.org/10.1103/PhysRevLett.103.211801}{Phys.\
  Rev.\ Lett.\  {\bf 103} (2009) 211801},
  \href{http://arxiv.org/abs/0807.4544}{{\tt arXiv:0807.4544}}\relax
\mciteBstWouldAddEndPuncttrue
\mciteSetBstMidEndSepPunct{\mcitedefaultmidpunct}
{\mcitedefaultendpunct}{\mcitedefaultseppunct}\relax
\EndOfBibitem
\bibitem{Aubert:2009ai}
\babar collaboration, B.~Aubert {\em et~al.},
  \ifthenelse{\boolean{articletitles}}{{\it {Measurement of \Dz--\Dzb mixing
  using the ratio of lifetimes for the decays $D^0\to K^-\pi^+$ and $K^+K^-$}},
  }{}\href{http://dx.doi.org/10.1103/PhysRevD.80.071103}{Phys.\ Rev.\  {\bf D
  80} (2009) 071103}, \href{http://arxiv.org/abs/0908.0761}{{\tt
  arXiv:0908.0761}}\relax
\mciteBstWouldAddEndPuncttrue
\mciteSetBstMidEndSepPunct{\mcitedefaultmidpunct}
{\mcitedefaultendpunct}{\mcitedefaultseppunct}\relax
\EndOfBibitem
\bibitem{delAmoSanchez:2010xz}
BaBar collaboration, P.~del Amo~Sanchez {\em et~al.},
  \ifthenelse{\boolean{articletitles}}{{\it {Measurement of \Dz--\Dzb mixing
  parameters using $\Dz\to\KS\pi^+\pi^-$ and $\Dz\to\KS K^+K^-$ decays}},
  }{}\href{http://dx.doi.org/10.1103/PhysRevLett.105.081803}{Phys.\ Rev.\
  Lett.\  {\bf 105} (2010) 081803}, \href{http://arxiv.org/abs/1004.5053}{{\tt
  arXiv:1004.5053}}\relax
\mciteBstWouldAddEndPuncttrue
\mciteSetBstMidEndSepPunct{\mcitedefaultmidpunct}
{\mcitedefaultendpunct}{\mcitedefaultseppunct}\relax
\EndOfBibitem
\bibitem{Asner:2012xb}
CLEO collaboration, D.~Asner {\em et~al.},
  \ifthenelse{\boolean{articletitles}}{{\it {Updated measurement of the strong
  phase in $\Dz\to K^+\pi^-$ decay using quantum correlations in $e^+e^- \to
  \Dz\Dzb$ at CLEO}},
  }{}\href{http://dx.doi.org/10.1103/PhysRevD.86.112001}{Phys.\ Rev.\  {\bf D
  86} (2012) 112001}, \href{http://arxiv.org/abs/1210.0939}{{\tt
  arXiv:1210.0939}}\relax
\mciteBstWouldAddEndPuncttrue
\mciteSetBstMidEndSepPunct{\mcitedefaultmidpunct}
{\mcitedefaultendpunct}{\mcitedefaultseppunct}\relax
\EndOfBibitem
\bibitem{LHCb-PAPER-2012-038}
LHCb collaboration, R.~Aaij {\em et~al.},
  \ifthenelse{\boolean{articletitles}}{{\it {Observation of \Dz--\Dzb
  oscillations}},
  }{}\href{http://dx.doi.org/10.1103/PhysRevLett.110.101802}{Phys.\ Rev.\
  Lett.\  {\bf 110} (2013) 101802}, \href{http://arxiv.org/abs/1211.1230}{{\tt
  arXiv:1211.1230}}\relax
\mciteBstWouldAddEndPuncttrue
\mciteSetBstMidEndSepPunct{\mcitedefaultmidpunct}
{\mcitedefaultendpunct}{\mcitedefaultseppunct}\relax
\EndOfBibitem
\bibitem{Aaltonen:2013pja}
CDF collaboration, T.~Aaltonen {\em et~al.},
  \ifthenelse{\boolean{articletitles}}{{\it {Observation of \Dz--\Dzb mixing
  using the CDF II detector}},
  }{}\href{http://dx.doi.org/10.1103/PhysRevLett.111.231802}{Phys.\ Rev.\
  Lett.\  {\bf 111} (2013) 231802}, \href{http://arxiv.org/abs/1309.4078}{{\tt
  arXiv:1309.4078}}\relax
\mciteBstWouldAddEndPuncttrue
\mciteSetBstMidEndSepPunct{\mcitedefaultmidpunct}
{\mcitedefaultendpunct}{\mcitedefaultseppunct}\relax
\EndOfBibitem
\bibitem{Alves:2008zz}
LHCb collaboration, A.~A. Alves~Jr. {\em et~al.},
  \ifthenelse{\boolean{articletitles}}{{\it {The \lhcb detector at the LHC}},
  }{}\href{http://dx.doi.org/10.1088/1748-0221/3/08/S08005}{JINST {\bf 3}
  (2008) S08005}\relax
\mciteBstWouldAddEndPuncttrue
\mciteSetBstMidEndSepPunct{\mcitedefaultmidpunct}
{\mcitedefaultendpunct}{\mcitedefaultseppunct}\relax
\EndOfBibitem
\bibitem{LHCb-DP-2012-003}
M.~Adinolfi {\em et~al.}, \ifthenelse{\boolean{articletitles}}{{\it
  {Performance of the \lhcb RICH detector at the LHC}},
  }{}\href{http://dx.doi.org/10.1140/epjc/s10052-013-2431-9}{Eur.\ Phys.\ J.\
  {\bf C 73} (2013) 2431}, \href{http://arxiv.org/abs/1211.6759}{{\tt
  arXiv:1211.6759}}\relax
\mciteBstWouldAddEndPuncttrue
\mciteSetBstMidEndSepPunct{\mcitedefaultmidpunct}
{\mcitedefaultendpunct}{\mcitedefaultseppunct}\relax
\EndOfBibitem
\bibitem{LHCb-DP-2012-004}
R.~Aaij {\em et~al.}, \ifthenelse{\boolean{articletitles}}{{\it {The \lhcb
  trigger and its performance in 2011}},
  }{}\href{http://dx.doi.org/10.1088/1748-0221/8/04/P04022}{JINST {\bf 8}
  (2013) P04022}, \href{http://arxiv.org/abs/1211.3055}{{\tt
  arXiv:1211.3055}}\relax
\mciteBstWouldAddEndPuncttrue
\mciteSetBstMidEndSepPunct{\mcitedefaultmidpunct}
{\mcitedefaultendpunct}{\mcitedefaultseppunct}\relax
\EndOfBibitem
\bibitem{PDG2012}
Particle Data Group, J.~Beringer {\em et~al.},
  \ifthenelse{\boolean{articletitles}}{{\it {\href{http://pdg.lbl.gov/}{Review
  of particle physics}}},
  }{}\href{http://dx.doi.org/10.1103/PhysRevD.86.010001}{Phys.\ Rev.\  {\bf D
  86} (2012) 010001}, {and 2013 partial update for the 2014 edition}\relax
\mciteBstWouldAddEndPuncttrue
\mciteSetBstMidEndSepPunct{\mcitedefaultmidpunct}
{\mcitedefaultendpunct}{\mcitedefaultseppunct}\relax
\EndOfBibitem
\bibitem{LHCb-PAPER-2012-026}
LHCb collaboration, R.~Aaij {\em et~al.},
  \ifthenelse{\boolean{articletitles}}{{\it {Measurement of the $D^\pm$
  production asymmetry in $7\tev$ $pp$ collisions}},
  }{}\href{http://dx.doi.org/10.1016/j.physletb.2012.11.038}{Phys.\ Lett.\
  {\bf B 718} (2013) 902–909}, \href{http://arxiv.org/abs/1210.4112}{{\tt
  arXiv:1210.4112}}\relax
\mciteBstWouldAddEndPuncttrue
\mciteSetBstMidEndSepPunct{\mcitedefaultmidpunct}
{\mcitedefaultendpunct}{\mcitedefaultseppunct}\relax
\EndOfBibitem
\bibitem{Grossman:2009mn}
Y.~Grossman, Y.~Nir, and G.~Perez, \ifthenelse{\boolean{articletitles}}{{\it
  {Testing new indirect CP violation}},
  }{}\href{http://dx.doi.org/10.1103/PhysRevLett.103.071602}{Phys.\ Rev.\
  Lett.\  {\bf 103} (2009) 071602}, \href{http://arxiv.org/abs/0904.0305}{{\tt
  arXiv:0904.0305}}\relax
\mciteBstWouldAddEndPuncttrue
\mciteSetBstMidEndSepPunct{\mcitedefaultmidpunct}
{\mcitedefaultendpunct}{\mcitedefaultseppunct}\relax
\EndOfBibitem
\bibitem{Kagan:2009gb}
A.~L. Kagan and M.~D. Sokoloff, \ifthenelse{\boolean{articletitles}}{{\it {On
  indirect CP violation and implications for \Dz--\Dzb and \Bs--\Bsb mixing}},
  }{}\href{http://dx.doi.org/10.1103/PhysRevD.80.076008}{Phys.\ Rev.\  {\bf D
  80} (2009) 076008}, \href{http://arxiv.org/abs/0907.3917}{{\tt
  arXiv:0907.3917}}\relax
\mciteBstWouldAddEndPuncttrue
\mciteSetBstMidEndSepPunct{\mcitedefaultmidpunct}
{\mcitedefaultendpunct}{\mcitedefaultseppunct}\relax
\EndOfBibitem
\end{mcitethebibliography}
